\documentclass[twocolumn,aps,prd,epsf,nofootinbib]{revtex4-1}

\usepackage{color}
\usepackage{amsmath}
\usepackage{amssymb}
\usepackage{braket}
\usepackage[T1]{fontenc}
\usepackage{url}
\usepackage[english]{babel}
\usepackage{graphicx}
\usepackage{subfigure}
\usepackage{graphicx}% Include figure files
\usepackage{dcolumn}% Align table columns on decimal point
\usepackage{bm}% bold math
\usepackage{float}
\usepackage{url}

\usepackage{bclogo}

\usepackage{tikz}
\usetikzlibrary{trees}
\usetikzlibrary{decorations.pathmorphing}
\usetikzlibrary{decorations.markings}

\usepackage{amsfonts} 

\usepackage{grffile}

\newcommand{\be}{\begin{equation}}
\newcommand{\ee}{\end{equation}}
\newcommand{\bea}{\begin{eqnarray}}
\newcommand{\eea}{\end{eqnarray}}

\newcommand{\bit}{\begin{itemize}}
\newcommand{\eit}{\end{itemize}}

\usepackage{verbatim}

\extrafloats{1000}

\graphicspath{{figs/}}

\begin{document}

%===================================================================================
\title{The lattice Landau gauge photon propagator for 4D compact QED}

\author{Lee C. Loveridge$^{1,2}$}
\email{loverilc@piercecollege.edu}
\author{Orlando Oliveira$^1$}
\email{orlando@uc.pt}
\author{Paulo J. Silva$^1$}
\email{psilva@uc.pt}

\affiliation{ \mbox{} \\ 
                 $^1$ CFisUC, Department of Physics, University of Coimbra, 3004-516 Coimbra, Portugal \\ \\
                 $^2$ Los Angeles Pierce College, 6201 Winnetka Ave.,  Woodland Hills CA 91371, USA}

%%%%%%%%%%%%%%%%%%%%%%%%%    abstract   %%%%%%%%%%%%%%%%%%%%%%%%%%%%%%%%%
\begin{abstract}
In this work we report on the Landau gauge photon propagator computed for pure gauge
4D compact QED in the confined and deconfined phases
and for large lattices volumes: $32^4$, $48^4$ and $96^4$. 
In the confined phase, compact QED develops mass scales that render the propagator finite at all momentum scales and
no volume dependence is observed for the simulations performed. Furthermore,
for the confined phase the propagator is compatible with a Yukawa massive type functional form. 
For the deconfined phase the photon propagator seems to approach a free field propagator as the lattice volume is increased. 
In both cases, we also investigate the static potential and the average value of the number of Dirac strings
in the gauge configurations $m$. 
In the confined phase the mass gap translates into a linearly growing static potential, while in the deconfined
phase the static potential approaches a constant at large separations.
Results shows that $m$ is, at least, one order of magnitude larger in the confined phase and confirm that the 
appearance of a confined phase is connected with the topology of the gauge group.
\end{abstract}

%\pacs{}
\maketitle

%\tableofcontents

%===================================================================================
%===================================================================================
\section{Introduction and Motivation}

The quest to understand quark confinement has, long ago, lead to the formulation of QED on a hypercubic lattice
 \cite{Wilson:1974sk} by Wilson. He was able to show that the compact formulation of QED has two phases. In the
 strong coupling limit, i.e. at low $\beta = 1/e^2$ where $e$ is the bare coupling constant, the theory is confining in the
 sense that the static potential between fermions grows linearly with the distance. In the weak coupling regime,
 i.e. at large $\beta$ values, it was argued that the results of perturbation theory for the continuum formulation
 should be recovered. These predictions for the phase diagram of QED have been confirmed with
 numerical simulations and by various theoretical analyses of the theory
 \cite{Banks:1977cc,Glimm:1977gz,Fradkin:1978th,Creutz:1979zg,Guth:1979gz,Lautrup:1980xr,Frohlich:1982gf,
 Berg:1983is,Kogut:1987cd,Jersak:1996mj,Gubarev:2000eu,Panero:2005iu}. 
 In numerical calculations, they typically use a mapping into its dual formulation associated 
with the $ \mathbb{Z}_n$ symmetry group  \cite{Balian:1974ts,Drouffe:1978xz} instead of performing a sampling of the compact formulation of QED directly. 

For compact QED the transition between the confining and the non-confining phase occurs at $\beta = \beta_c \approx 1$, 
see \cite{Lautrup:1980xr,Arnold:2002jk} and references therein. For $\beta < \beta_c$ the static potential $V(R)$ is compatible with a Cornell type of potential
\begin{equation}
   V(R) = V_0 + \frac{a}{R} + \sigma \, R \ ,
\end{equation}
with the dimensionless string tension $\sigma$ being a decreasing function of $\beta$ as its critical value is approached from below \cite{Panero:2005iu}.
Numerical simulations for $\beta > \beta_c$ suggest that the theory contains a massless photon and reproduces the behaviour associated with a free field theory.
However, in the numerical simulations the reproduction of a free field theory for $\beta > \beta_c$ is 
never perfect \cite{Berg:1983is,Coddington:1987yz,Nakamura:1991ww}.

Compact QED, having a confined and a deconfined phase, can also be viewed as a laboratory to try to understand the differences between the two
phases. In 3D, the confining mechanism is associated with the presence of monopole configurations \cite{Polyakov:1975rs,Polyakov:1976fu} at low $\beta$ 
values; see also \cite{Chernodub:2002gp,Bakker:2000kg} for numerical simulations. 
However, in 4D these classical configurations or their equivalent are not able to generate a linearly rising potential at large separations and, therefore,
cannot be at the origin of the confining mechanism. It has been argued that a Cornell potential type for the static potential is related to the topology of the gauge group,
see e.g. \cite{Greensite:2011zz} and references therein. 

The interest in compact QED is not limited to its phase diagram. For example, it is not yet clear if the continuum limit of compact QED is a sensible theory. 
If the fermion sector of the theory is ignored, its continuum theory is a free field theory that one expects to recover for sufficiently high $\beta$ values. 
The simulations performed so far suggest that this is the case, but full agreement with the results of a free field theory has not been achieved. 
The transition of the confined phase to the deconfined phase seems to be first order which, once more, raises the question of how to take the continuum limit
for compact QED. 
Besides the question of the continuum limit, $U(1)$ gauge theories are relevant to understand the Standard Model, the comprehension of the Higgs sector calls for 
simulations of Abelian Higgs models, and to describe many properties in Condensed Matter Physics. 
Furthermore, recently the lattice QCD community has started to compute QED corrections to QCD and, certainly, 
a good understanding of lattice Abelian models is of paramount importance. $U(1)$ gauge theories are also a laboratory towards building simulations
of gauge theories on quantum computers.

In the current work we are mainly interested in identifying the differences between the confined and deconfined phase by looking at the Landau gauge
photon propagator in momentum space for the two phases of compact QED in the quenched approximation, i.e. for the compact U(1) theory. 
The simulations described in this work do not take into account the contribution
of the fermionic degrees of freedom to the dynamics.
For the confined phase the theory develops a mass scale and the photon 
propagator is finite and, at least for the simulations discussed, the presence of the mass scale prevents or reduces the finite volume effects.
On the other hand, in the deconfined phase the photon propagator is compatible with a divergent $1/k^2$ or a higher power of $k$ behaviour at low momenta 
$k$. Finite volume effects are sizeable and the propagator approaches that of a free field theory as the lattice volume is increased.
We hope that by studying the photon propagator one can arrive at a better understanding of the confinement mechanism for QCD.

The Landau gauge photon propagator for the confined phase qualitatively follows the same type of behaviour as the Landau gauge 
gluon propagator in QCD for pure Yang-Mills theories; see, for example,
\cite{Cucchieri:2007md,Bogolubsky:2009dc,Dudal:2010tf,Oliveira:2012eh,Duarte:2016iko,Cucchieri:2016jwg,Dudal:2018cli} 
for lattice simulations of the Landau gauge gluon propagator and
\cite{Huber:2020keu,Huber:2018ned,Cyrol:2016tym,Fischer:2008uz,Aguilar:2008xm,Boucaud:2008ky,Dudal:2008sp,Alkofer:2003jj}
for continuum estimations of the same correlation function (see also the references therein).
This suggests that the confinement mechanisms for the gluon and the photon have some similarities and, in particular, that the confining
theory is associated with dynamically generated mass scales that make the propagator finite in the full momentum range for compact QED and for QCD. 
For QED a photon mass has been related to a meaningful physical value of the
expectation value of the vector potential squared, that is connected with the existence of topological structures for the theory \cite{Gubarev:2000eu}.

Our work  also aims to fill the gap in the literature \cite{Nakamura:1990dq,Nakamura:1991ww}
and provide a large volume 4D lattice simulation of the Landau gauge photon propagator in momentum space for $\beta$ below and above $\beta_c$.
The two phases are distinguished not only by the qualitative behaviour of the propagators but also
 by their topological structures and the static potential as computed from the Wilson loops.
Our findings for the topological structures and static potential confirm previous results that can be found in the literature.

This work is organised as follows. In Sec. \ref{Sec:setup} we provide the definitions used in our calculation ranging from the Wilson action,
the gauge fixing procedure, the definition of the electromagnetic potential, the propagator and number of Dirac strings crossing each plaquette.
In Sec. \ref{Sec:confined} the propagator and the static potential are discussed for the confined phase, while in
Sec. \ref{Sec:DeconfPhase} we discuss these properties for the deconfined phase. Finally, in Sec. \ref{sec:fim} we summarise our results
and discuss the differences between the two phases.

%===================================================================================
%===================================================================================
\section{Compact QED: definitions, lattice setup and details of the simulation \label{Sec:setup}}

In the current work we will consider the compact version of QED defined on an hypercubic lattice that is 
described by the Wilson action which, in Euclidean space, is given by
\begin{equation}
   S_W (U) = \beta \sum_x \sum_{1 \leqslant \mu, \nu \leqslant 4} \left\{ 1 - \Re \, \left[ U_{\mu\nu} (x) \right]\right\} \, ,
   \label{Eq:accao}
\end{equation}
where $\beta = 1/ e^2$, with $e$ being the bare coupling constant, and the plaquette operator
\begin{equation}
   U_{\mu\nu} (x) = U_\mu (x) \, U_\nu (x + a \, \hat{e}_\mu) \, U^\dagger_\mu (x + a \, \hat{e}_\nu) \,  U^\dagger_\nu (x)
   \label{Eq:plaquette}
\end{equation}
is written in terms of the link fields
\begin{equation}
   U_\mu (x) = \exp\left\{ i \, e \, a\, A_\mu \left( x + \frac{a}{2} \, \hat{e}_\mu \right) \right\} \ ,
   \label{Eq:Link}
\end{equation}
with $a$ being the lattice spacing, $\hat{e}_\mu$ the unit vector along direction $\mu$ and $A_\mu$ is the bare photon field. 
In the continuum limit the plaquette operator (\ref{Eq:plaquette}) can also be written 
\begin{equation}
   U_{\mu\nu} (x) = \exp\left\{ i \, e \, \oint_C A_\mu (z)  \, dz_\mu \right\}
\end{equation}
where $C$ is any closed curve around point $x$ and infinitesimally close to it. 
On an hypercubic lattice, the term that appears in the exponential is the change of the photon field 
around a plaquette centered at $x + a \, ( \hat{e}_\mu + \hat{e}_\nu ) /2$ and we write
\begin{equation}
   U_{\mu\nu} (x) = \exp\left\{ i \, e \, a \, \Big( \, \Delta A_{\mu\nu} (x) \, \Big)  \right\}
\end{equation}
It follows from the definitions that everywhere in the lattice $- \pi \leqslant e \, a\, A_\mu \leqslant \pi$ and
$- \pi \leqslant e \, a ~ \Delta A_{\mu\nu} (x) \leqslant \pi$, i.e. the quantities $e \, a\, A_\mu$ and  $e \, a ~\Delta A_{\mu\nu}$  take values on compact spaces.
Note that, in general, $e \, a ~\Delta A_{\mu\nu}$ is not given by the sum of $e \, a\, A_\mu$ over each of the links that define the plaquette but, instead,
\begin{eqnarray}
\Delta A_{\mu\nu} (x) & = &
 A_\mu \left( x + \frac{a}{2} \, \hat{e}_\mu \right) +
                    A_\nu \left( x + a \, \hat{e}_\mu + \frac{a}{2} \, \hat{e}_\nu \right) \nonumber \\
                    & & \quad -
                    A_\mu \left( x +  a \, \hat{e}_\nu + \frac{a}{2} \, \hat{e}_\mu  \right) -
                   A_\mu \left( x + \frac{a}{2} \, \hat{e}_\nu \right)   \nonumber \\
                   & & \qquad\quad + \, \frac{ 2 \, \pi \, m_{\mu\nu} (x)}{e \, a }
                   \label{EqDirac_string0}
\end{eqnarray}
where $m_{\mu\nu}(x)$ is an integer number that measures the number of Dirac strings that cross the 
plaquette\footnote{In 3D, $m_{\mu\nu}(x)$ can be identified with the number of monopoles in the plaquette.}.
The integer field $m_{\mu\nu}(x)$ can be measured by combining information on the links and plaquettes \cite{DeGrand:1980eq}. 
We will not study  $m_{\mu\nu}(x)$ in detail but will report on its mean value over the lattice\footnote{Note that $m$ is
the mean value of $| m_{\mu\nu}(x) |$ and not of $m_{\mu\nu}(x)$. The latter quantity is gauge invariant, while the first is not. The rationale for
using the absolute value instead of $m_{\mu\nu}(x)$ being that the mean value of $m_{\mu\nu}(x)$ is quite a small number
since the number of positive and negative values of $m_{\mu\nu}(x)$ is roughly equal.}, i.e.
\begin{equation}
 m = \frac{1}{6 \, V} \, \sum_{x, \mu <\nu}  | m_{\mu\nu}(x) |
 \label{Eq:mean_dirac_string}
\end{equation}
where $V$ is the number of lattice points. Indeed, as discussed below $m$ can be used to distinguish between the confined and deconfined phases,
with the configurations in the confined phase having a much larger mean number of  Dirac strings crossing each plaquette.

The generating functional of the compact QED Green's functions is given by
\begin{equation}
   Z = \int \, \mathcal{D} A ~ \exp\left\{ - S_W(U) \right\} \ ,
   \label{Eq:Z}
\end{equation}
where $S_W$ is defined in Eq. (\ref{Eq:accao}).
For the importance sampling we rely on an implementation of the Hybrid Monte Carlo method \cite{Duane:1987de} based on QDP++ and Chroma libraries \cite{Edwards:2004sx}. 

The rotation of the links obtained by importance sampling to the Landau gauge is formulated as an optimization problem, over the gauge orbits.
Setting the optimization problem depends on the definition of the photon field. 

The gauge field can be computed with a linear definition that, 
for the $U(1)$ theory, reads
\begin{equation}
   e \, a \, A_\mu \left( x + \frac{a}{2} \hat{e}_\mu \right)  = \frac{U_\mu(x) - U^\dagger_\mu(x)}{2 \, i} \ .
   \label{Eq:GaugeFieldLinear}
\end{equation}
If one relies on this definition the gauge fixing is performed by maximizing the functional
\begin{equation}
   F[U; g] = \frac{1}{V \, D} \sum_{x,\mu} \, \Re \left[  \,  g(x) \, U_\mu (x) \, g^\dagger(x + a \, \hat{e}_\mu )\,  \right] \ ,
   \label{Eq:Landau_linear}
\end{equation}   
where $V$ is the total number of lattice points and $D$ the Euclidean spacetime dimension. It can be shown, see e.g.  \cite{Silva:2004bv},
that in this way the continuum Landau gauge condition is reproduced up to corrections $\mathcal{O}(a^2)$.
For the $U(1)$ gauge theory there is no clear way to set the lattice spacing and, therefore, this procedure can introduce significant 
deviations of the continuum Landau gauge when applied to the confined ($\beta \lesssim 1$) or to the deconfined ($\beta \gtrsim 1$) phase. 
The setup just described, that uses a linear definition of the gauge field, is used for non-Abelian gauge theories defined on a lattice from simulations that
are close to continuum physics.  In a first step to define the Landau gauge for the photon field, we used the above procedure relying on a steepest descent 
algorithm with Fourier acceleration, see  \cite{Silva:2004bv} for details and references, and controlling the approach to the Landau gauge with the quantity
\begin{equation}
   \Delta (x) = \sum_\nu \Big[ U_\nu ( x - a \, \hat{e}_\nu) -  U_\nu ( x ) \Big] \ ,
\end{equation}
a lattice version of $- \, \partial \cdot A(x)$. The maximisation was stopped when
\begin{equation}
  \theta = \frac{1}{V} \sum_x \Big| \Delta (x) \Big|^2  \,  <  \, 10^{-15} \ .
  \label{Eq:Theta}
\end{equation}

On the other hand, the Euclidean photon field can be computed using a logarithmic definition
\begin{equation}
   e \, a \, A_\mu \left( x + \frac{a}{2} \hat{e}_\mu \right) = -i \, \ln \Big(  U_\mu (x) \Big) \ .
   \label{Eq:photon_field}
\end{equation} 
This is an exact definition, up to machine precision, that does not call for the use of a small lattice spacing.
Then, following \cite{Ilgenfritz:2010gu} adapted to the Abelian theory, the Landau gauge condition is achieved by maximizing the functional
\begin{equation}
   \widetilde{F}[U;g] = \frac{1 }{V \, D } \sum_{x ,\mu} \,  \bigg\{ 1 -  a^2 e^2 \left[ A^{(g)} _\mu \left( x + \frac{a}{2} \hat{e}_\mu \right) \right]^2 \bigg\}
   \label{Eq:Landau_Log}
\end{equation}
over the gauge orbits. In the Eq. (\ref{Eq:Landau_Log}) the field $e \, a \, A^{(g)}$ is the photon field given by
Eq. (\ref{Eq:photon_field}) after the links $U_\mu(x)$ have been gauge transformed by $g(x)$. The approach towards the Landau gauge can be monitored using
\begin{equation}
   \widetilde\Delta (x) = a\, e \, \sum_\nu \left[ A_\nu ( x - \frac{a}{2} \, \hat{e}_\nu) -  A_\nu ( x + \frac{a}{2} \, \hat{e}_\nu ) \right] \ ,
\end{equation}
once more a lattice version of $- \, \partial \cdot A(x)$. In our computation of the Landau gauge propagator, after the maximization problem associated
with the functional given in Eq. (\ref{Eq:Landau_linear}), a maximization of the functional (\ref{Eq:Landau_Log}) is also performed. In this way one aims to
reduce possible deviations of the continuum Landau gauge for both phases of the theory.
In this second maximization we use again a steepest descent algorithm with Fourier acceleration and the gauge fixing was stopped when
\begin{equation}
   \widetilde{\theta} = \frac{1}{V} \sum_x \left| \widetilde\Delta (x)  \right|^2 < 10^{-15} \ .
\end{equation}

The definition of gauge field potential from the link variables, its implications on the choice of the optimization functional to define
the Landau gauge on the lattice, how the different definitions impact the gluon correlation functions as been the subject of several studies, see e.g.
\cite{Giusti:1998ur,Cucchieri:1998ta,Cucchieri:1999dt,Nakajima:2000kh,Bogolubsky:2002ui,Catumba:2021hcx}
and references therein. These studies address the problem for non-Abelian gauge theories and explore essentially the linear definition
given in Eq. (\ref{Eq:GaugeFieldLinear}) and its variants. For the non-Abelian gauge theories the different choices for the definition of the gauge
potential lead to the same propagators. However, as already stated, the linear definition is possible due to asymptotic freedom of $SU(N)$ gauge
theories. For a $U(1)$ gauge theory on the lattice there is not a clear way of setting the physical scale and, therefore, to define the lattice spacing
that would allow (or not) to the use for the linear definition of the photon field.

In both stages the maximisation of the gauge fixing functional is done with a Fourier accelerated steepest descent method that calls for
the PFFT library \cite{Pippig2013} to do the required fast Fourier transformations. 
The complete numerical simulation, i.e. the importance sampling, the gauge fixing and the computation of all quantities, 
were performed in the Navigator cluster \cite{Navigator} of the University of Coimbra.

The use of the Hybrid Monte Carlo to perform the importance sampling can be questionable as, long ago
\cite{Grosch:1985cz,Kerler:1995va}, very long-living metastable states were observed in the Monte Carlo history of the system.
In our study, the Monte Carlo time evolution of the mean value of the plaquette does not reveal metastable states. Indeed, at least for the two
values of $\beta$ considered, that are far from the transition between the two known phases,  the Monte Carlo history for the mean 
value of the plaquette show a initial thermalization phase followed by a fluctuation pattern. 
No sign of metastable states are found and we have considered large Monte Carlo time histories. 
The time histories are larger that those mentioned in \cite{Grosch:1985cz} but our lattice sizes are much larger than those considered there.
However, the good agreement %...., at least quantitative, 
between all the simulations for each of the $\beta$ values and the non-observation
of metastable states 
%Furthermore, we did check that the thermalized values of the fluctuation pattern for the mean value of the plaquette do
%not depend on the initial configuration where the Monte Carlo is started. Although this is no proof that the metastable states are not there but
%it 
gives confidence on the results of the simulation.

Another issue of concern connected with the computation of the photon propagator is the presence of Gribov copies. In the lattice
formulation of the compact $U(1)$ gauge theory, the Gribov copies are associated with different extrema of the maximizating functional 
that defines the lattice Landau gauge, see Eqs (\ref{Eq:Landau_linear}) and (\ref{Eq:Landau_Log}). 
The presence of the Gribov in the Landau gauge was also studied long ago, see e.g. 
\cite{Bornyakov:1993yy,Bogolubsky:1999cb,Bogolubsky:1999dv}, with observed measurable effects in the photon propagator. 
There are two important differences between our implementation of the Landau gauge and that used in the those earlier studies. 
The first one is the precision on the definition of the lattice version of $\partial A = 0$ that differs by several orders of
magnitude. It well known that from the analysis of the non-Abelian theories that a small value of $\theta$ of $10^{-12}$ or smaller
should be used to have a good implementation of the Landau gauge on the lattice. 
The relaxation of $\theta$ towards higher values affects mainly the low momentum region 
propagator\footnote{The optimizing functionals are equivalent to look at the extrema of $\int d^4x \, ||A^{g}||^2$, where
$A^{(g)}$ is the gauge transformed field $A$, along the gauge orbits that in momentum space becomes
$\int dk d \Omega \, k^3  \,  [A^{g}(k) ]^2$, where $d \Omega$ stands for the angular integration. It follows that
the lower momenta are more sensitive to the quality of the gauge fixing.}. The second
difference is related to the definition of the gauge field. The previous studies use the linear definition for 
the gauge field (\ref{Eq:GaugeFieldLinear}), while we use the logarithmic definition (\ref{Eq:photon_field}). This does not eliminate the
problem of the Gribov copies but it is not obvious that the observations reported in
\cite{Bornyakov:1993yy,Bogolubsky:1999cb,Bogolubsky:1999dv} apply in our formulation. The studies of pure gauge
non-Abelian gauge theories, see e.g. \cite{Silva:2004bv}, show that when $\theta \sim 10^{-12}$ or smaller and for statistics with
similar number of gauge configurations, typically, the effects of Gribov copies are diluted within the statistical error. We hope the same applies
to compact $U(1)$ theory and, in the following, we will assume that this is the case.

From the definition (\ref{Eq:photon_field}) for the Euclidean spacetime photon field, 
the momentum space photon field is given by
\begin{equation}
  A_\mu (p) =  \sum_x \, e^{ -i p \cdot( x + \frac{a}{2} \hat{e}_\mu )}  A_\mu \left( x + \frac{a}{2} \hat{e}_\mu \right) 
\end{equation}
and the Landau gauge propagator reads
\begin{equation}
   \langle A_\mu (p_1) ~ A_\mu(p_2) \rangle = V \, \delta( p_1 + p_2 ) \, D_{\mu\nu}(p_1)
   \label{Eq:prop1}
\end{equation}
where $\langle \cdots \rangle$ stands for the vacuum expectation value. In a lattice simulation, the vacuum expectation values are accessed 
via the generation of a set of configurations sampled accordingly with the probability distribution (\ref{Eq:Z}) and taking averages of the products of
gauge fields, 
such as those
in Eq. (\ref{Eq:prop1}), over the full set of gauge configurations. 
For the analysis of the propagator, it will be assumed that the propagator has the same tensor structure as the continuum theory, i.e.
\begin{equation}
D_{\mu\nu}(p) = \left( \delta_{\mu\nu} - \frac{p_\mu p_\nu}{p^2} \right) \, D(\hat{p})
\label{Eq:photon_prop}
\end{equation}
where the function $D(\hat{p})$, named propagator below, is a function of the tree level improved momenta
\begin{eqnarray}
  & &
    \hat{p} = \frac{2}{a} \sin \left( \frac{\pi}{L} \, n_\mu \right) \ , \nonumber \\
    & &  \quad\quad
     n_\mu = -\frac{L}{2}, \, -\frac{L}{2} + 1, \, \dots , \, 0, \, 1, \, \dots , \, \frac{L}{2}  - 1
\end{eqnarray}    
where $L$ is the number of lattice points in each side of the hypercubic lattice. The rationale to use $\hat{p}$ instead of the naive lattice momenta
\begin{equation}
    p =  \frac{2 \, \pi}{a \, L} \, n_\mu \ , 
\end{equation}    
comes from lattice perturbation theory that requires $\hat{p}$ instead of $p$. In the lattice evaluation of the gluon propagator the improved momentum
also helps to suppress  finite spacing effects in the propagator \cite{Leinweber:1998uu}.
In order to  further suppress the effects due to the use of finite lattice spacing we perform the conical and cylindrical cuts introduced in \cite{Leinweber:1998uu} for 
momenta $a \, \hat{p} > \Lambda_{\text{IR}}$ and, following the procedure devised in \cite{Dudal:2018cli}, below this threshold we consider all the momenta available to get information
on the infrared region. 
The choice of the infrared threshold $\Lambda_{\text{IR}}$ is a compromise between taking into account extra data, allowing larger fluctuations,
and resulting in a smooth curve for $D(p^2)$. The choice of this threshold does not change the overall behaviour of the lattice data and $\Lambda_{\text{IR}}$ will
be chosen differently for each simulation. In the following we use $\Lambda_{\text{IR}} = 0.2$ for the largest lattice volume and $\Lambda_{\text{IR}} = 0.4$
for the two smallest lattices.

The description of the lattice propagator with the continuum tensor structure as given by Eq. (\ref{Eq:photon_prop}) is questionable, especially concerning
the confined phase. Similar studies for the gluon propagator show that the lattice propagator has other tensor structures not considered in Eq. (\ref{Eq:photon_prop})
and, in principle, they should also be considered here. However, given that the definition of the lattice Landau gauge returns a transverse gauge field, one expects
a gauge propagator that should also be transverse. Furthermore, the studies performed for the gluon propagator suggest that the introduction of momentum cuts
selects the set of momenta where the finite lattice effects are minimised.
This gives us confidence that the same should apply to the photon propagator. 

If one assumes
a tensor structure as given by Eq. (\ref{Eq:photon_prop}), then the type of momentum considered in the projector is irrelevant as long as one measures the propagator
form factor using
\begin{equation}
   D(\hat{p}) = \frac{1}{3} \sum_{\mu = 1}^4 D_{\mu\mu} (p) \ .
\end{equation}
We remind the reader that for zero momentum the propagator is given by $\delta_{\mu\nu} D(0)$ and, therefore, the computation of $D(0)$ requires a different
normalisation factor.

In the current work, we aim to see how the photon propagator behaves in the confined and  deconfined phases. To achieve such a goal
we perform Monte Carlo simulations of the theory at $\beta = 0.8$ (confined phase) and at $\beta = 1.2$ (deconfined phase). 
In order to check for finite volume effects in both cases we perform simulations on $32^4$, $48^4$ and $96^4$ hypercubic lattices.
For each $\beta$ value and lattice volume, the propagators were computed using the last (in the Markov chain) 200 gauge configurations.
The configurations used in the calculation of propagator have a separation of 10 trajectories for the smaller lattice volume, for both $\beta$ values considered
herein, and also for the $48^4$ simulation in the confined phase ($\beta = 0.8$). In the remaining simulations we used a separation of 100 trajectories 
in the corresponding Markov chain.

\begin{figure}[t] %  figure placement: here, top, bottom, or page
   \centering
   \includegraphics[width=3.4in]{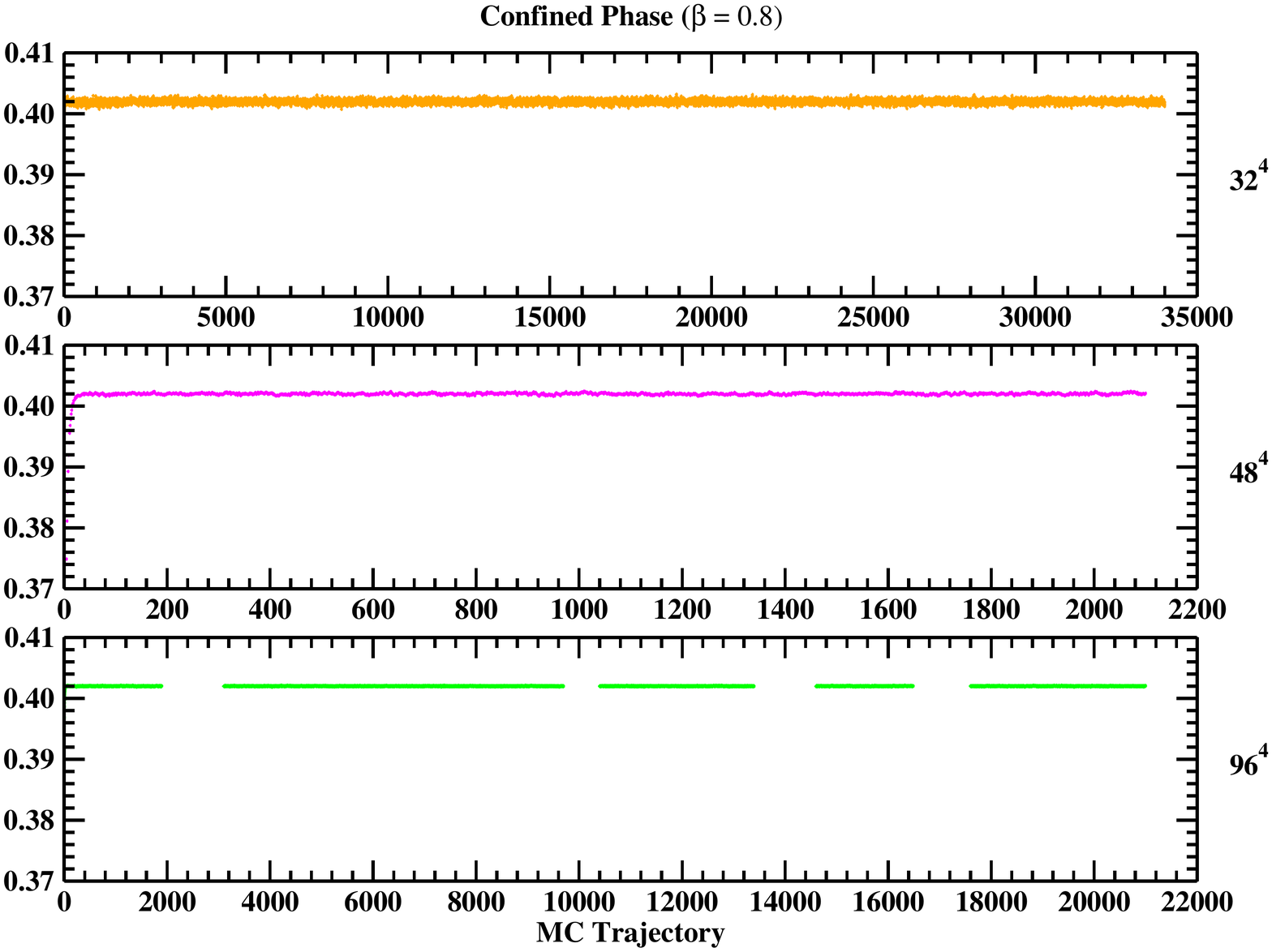}  \\
   \includegraphics[width=3.4in]{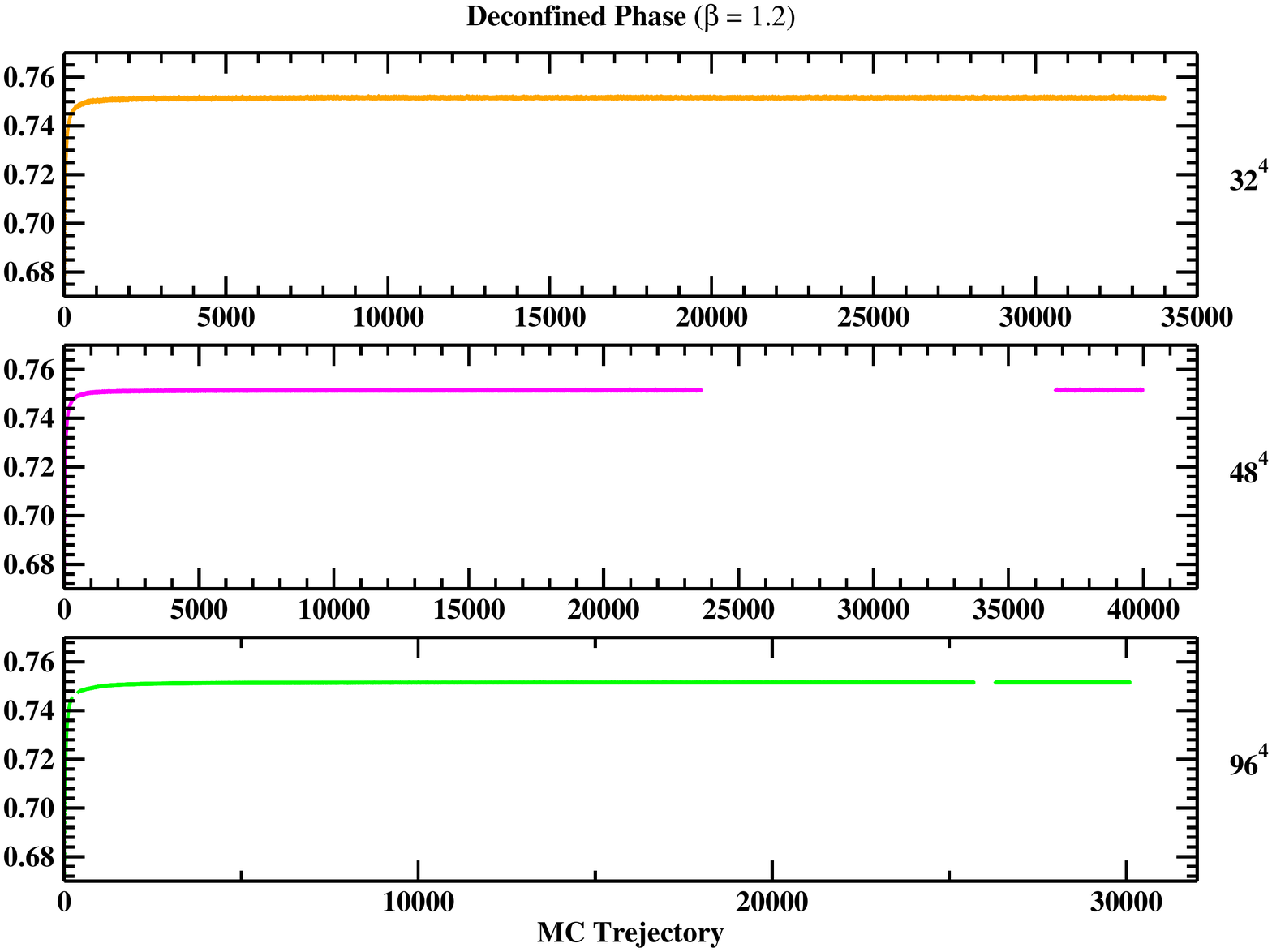} 
   \caption{Mean value, over the lattice, of the plaquette for all simulations. In the simulation we did not always keep the values of the plaquette for all computed trajectories and, therefore, the plots have regions with no data.}
   \label{fig:plaquettes}
\end{figure}

 In Fig. \ref{fig:plaquettes} the mean values of the plaquette over the lattice are shown for each of the Markov chains. 
In the simulations the value of the plaquette at the end of each trajectory was not always kept and, therefore, in the reconstruction of the plaquette history
we lost some of the data points. As the Fig. shows the mean value of the plaquette seems to be independent of the lattice volume for each $\beta$ and
in the deconfined region, i.e. for the simulation with $\beta = 1.2$, the plaquette is significantly larger. This result suggests that the $U(1)$
links approach unity as $\beta$ is increased.

For the computation of statistical errors for all the quantities reported here, i.e. propagators, Wilson loops and monopole densities,
we rely on the bootstrap method with a 67.5\% confidence level. The quoted errors associated with the fits assume Gaussian error propagation.

%===================================================================================
%===================================================================================
\section{Photon in the confined phase \label{Sec:confined}}

\begin{figure}[t] %  figure placement: here, top, bottom, or page
   \centering
   \includegraphics[width=3.4in]{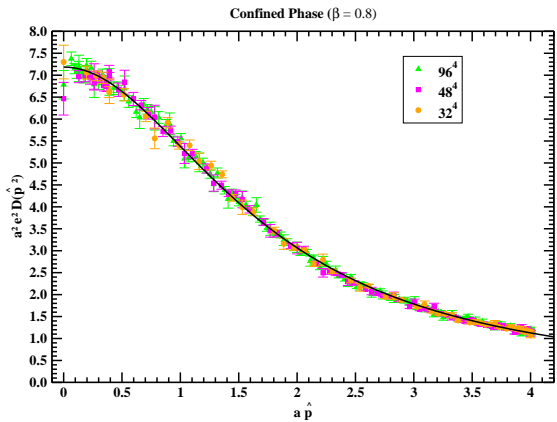}  
   \caption{The Landau photon propagator in the confined phase for all the lattice volumes. The solid black line refers to the fit to the $96^4$ lattice data discussed in the text. See text for further details.}
   \label{fig:photon_conf}
\end{figure}

The Landau gauge photon propagator for compact QED in the confined phase with $\beta = 0.8$
and for the various lattice volumes can be seen in Fig.
\ref{fig:photon_conf}. 
The data does not follow the behaviour of a free particle propagator and deviations from a $1/p^2$ functional form are clearly seen.
Indeed, the various data sets seem to be closer to the qualitatively behaviour of
the QCD gluon propagator \cite{Dudal:2018cli,Duarte:2016iko,Oliveira:2012eh}.
Moreover, the propagator being finite over the full range of momenta suggests that 4D compact QED generates  a mass gap dynamically,
as is also observed in 3D simulations \cite{Chernodub:2002gp,Bakker:2000kg}.
The data for the various volumes is compatible within one standard deviation and, therefore, shows no volume dependence. 

It seems that the presence of the 
mass gap is sufficient to reduce the volume dependence of $D(p^2)$. 
This contrasts with what is observed for the propagator in the deconfined phase; more on this topic later.

A possible way to identify the mass gap is by fitting the lattice data to a given functional form. We found that, for all volumes, the lattice data is
well described by a Yukawa type propagator
\begin{equation}
   a^2 e^2 D(p^2) = \frac{z_0}{p^2 + m^2} \ ,
   \label{Eq:confined_prop_fit}
\end{equation}
where $z_0$, $p^2$ and $m^2$ are dimensionless quantities.
The fits using the full range of momenta
result in a $\chi^2/d.o.f. =$ 1.57 for the data associated with the simulation using a $32^4$ lattice, 1.31 for the $48^4$ lattice data
and 1.08 for the  $96^4$ lattice data. The corresponding fitting parameters are
$z_0 = 21.37(14)$, $m^2  = 2.968(38)$,
$z_0 = 21.28(10)$, $m^2 = 2.963(24)$,
$z_0 = 21.399(66)$, $m^2 = 2.978(17)$, respectively, and are all compatible within one standard deviation.
We have observed that increasing $\Lambda_{\text{IR}}$ results in smaller values for the $\chi^2/d.o.f.$ in all cases.
In Fig. \ref{fig:photon_conf} the solid black line represents the functional form given in Eq. (\ref{Eq:confined_prop_fit}) with $z_0$ and $m^2$ given
by the estimation of the fit to the lattice data from the largest volume. Similar curves using the other two sets of parameters could be drawn but the curves are
indistinguishable to the naked eye from the curve shown.

\begin{figure}[t] %  figure placement: here, top, bottom, or page
   \centering
   \includegraphics[width=3.4in]{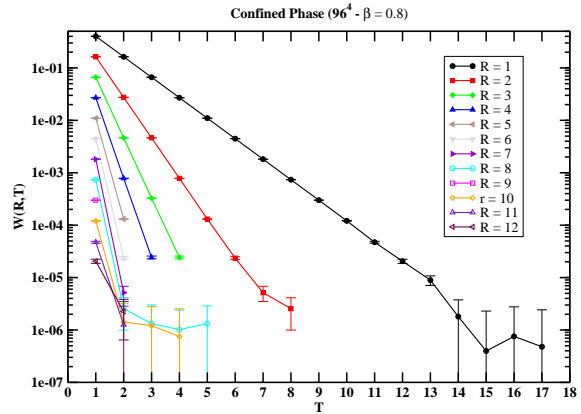}  
   \caption{Wilson loop $W(R,T)$ at $\beta = 0.8$ and for $L = 96$.}
   \label{fig:wilson_loop_conf}
\end{figure}

\begin{figure}[t] %  figure placement: here, top, bottom, or page
   \centering
   \includegraphics[width=3.4in]{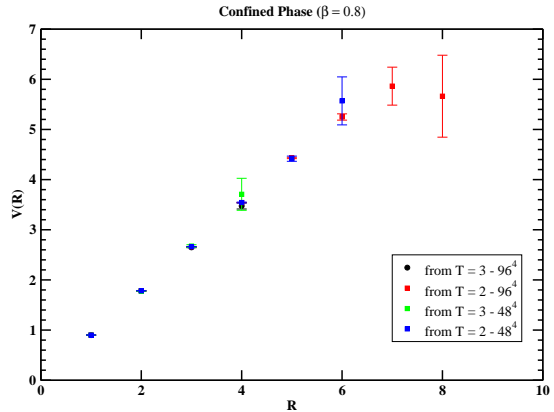}  
   \caption{The static potential $V(R)$ for $\beta = 0.8$ and for $L = 96$ and $L = 48$.}
   \label{fig:Vr_conf}
\end{figure}

The low $\beta$ phase of compact QED was investigated by Wilson in \cite{Wilson:1974sk}, where he computed the static potential from Wilson loops.
Indeed, it was shown that, at low $\beta$, the Wilson loop follows an area law and, therefore, the associated static potential grows linearly with the distance between
sources. It is in this sense that compact QED at low $\beta$ values is a confining theory. This observation motivated us to compute Wilson loops
and we took only those loops whose spatial part is along one of the lattice axis to measure the static potential $V(R)$. The Wilson loop can be seen in Fig.
\ref{fig:wilson_loop_conf}. 
Note that we use no trick to improve the signal to noise ratio, the noise for $W(R,T)$ is large for some cases and increases with $R$. This is an
indication that the static potential grows with $R$ given that
\begin{equation}
   W(R,T) = e^{-V(R) T} \ .
\end{equation}
Further, it is clear from Fig. \ref{fig:wilson_loop_conf} that  exponential behaviour sets in for quite small $T$. Then, from the data for $W(R,T)$ one can measure
$V(R)$ from
\begin{equation}
  V(R) = \log \left( \frac{W(R,T)}{W(R,T+1)} \right)
\end{equation}
and in Fig. \ref{fig:Vr_conf} we show $V(R)$ computed from taking $T = 2$ and $T = 3$. The data in Fig. \ref{fig:Vr_conf}  should be regarded as an upper bound
on $V(R)$. The results summarised in Figs. \ref{fig:wilson_loop_conf} and \ref{fig:Vr_conf} 
 confirms that $V(R)$ grows with $R$ and suggest that  the data is compatible with linear behaviour at large $R$.
In this sense the simulation confirms that compact QED is a confining theory at low $\beta$ values.

The static potential for 4D compact QED was computed in \cite{DeGrand:1981yq,Cella:1997hw,Panero:2005iu}
 from Polyakov loops, exploring duality transformations, 
and it was found that in the confined phase $V(R)$ grows linearly with the distance for sufficiently large $R$ as also found in our simulations.

%===================================================================================
%===================================================================================
\section{Photon in the deconfined phase \label{Sec:DeconfPhase}}

\begin{figure}[t] %  figure placement: here, top, bottom, or page
   \centering
   \includegraphics[width=3.4in]{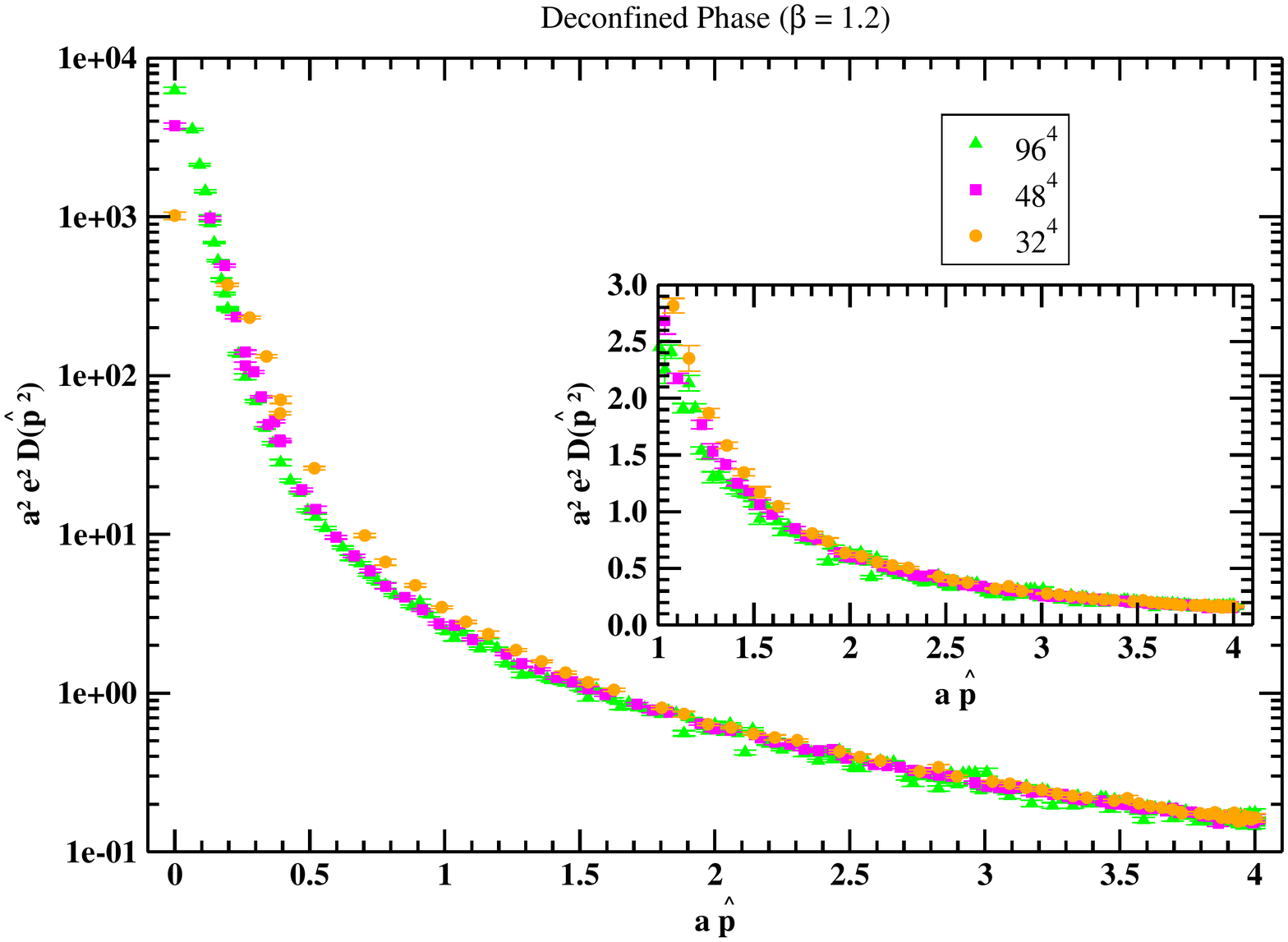}  
   \caption{The Landau photon propagator in the deconfined phase for all the lattice volumes. Note that the in the inner plot, the vertical scale is linear.}
   \label{fig:photon_deconf}
\end{figure}

The nature of the photon propagator at large $\beta$ is expected to be rather different than that observed in Fig. \ref{fig:photon_conf}. Indeed, as can be seen in Fig. 
\ref{fig:photon_deconf}, for the deconfined phase with $\beta = 1.2$
the photon propagator seems to diverge at zero momentum. Furthermore, if at low $\beta$ the propagator
is blind to the finite volume effects, the data for the various volumes in Fig. \ref{fig:photon_deconf} are not compatible with each other within one standard 
deviation. The propagator data for the smallest volume $32^4$ is above the other two sets of propagator data in the mid range momenta, 
while the data associated with the $48^4$ lattice is between the data computed with the smallest and the largest lattice volumes. 
However, at zero momenta the largest $a^2 D(0)$ is associated with the largest volume, followed by the $48^4$ data and by the
$32^4$ data in decreasing order of values. For momenta such that $a \, p \gtrsim 2$ all the data sets seems to be compatible within one standard deviation,
see the inner plot in Fig. \ref{fig:photon_deconf}.

\begin{figure}[t] %  figure placement: here, top, bottom, or page
   \centering
   \includegraphics[width=3.5in]{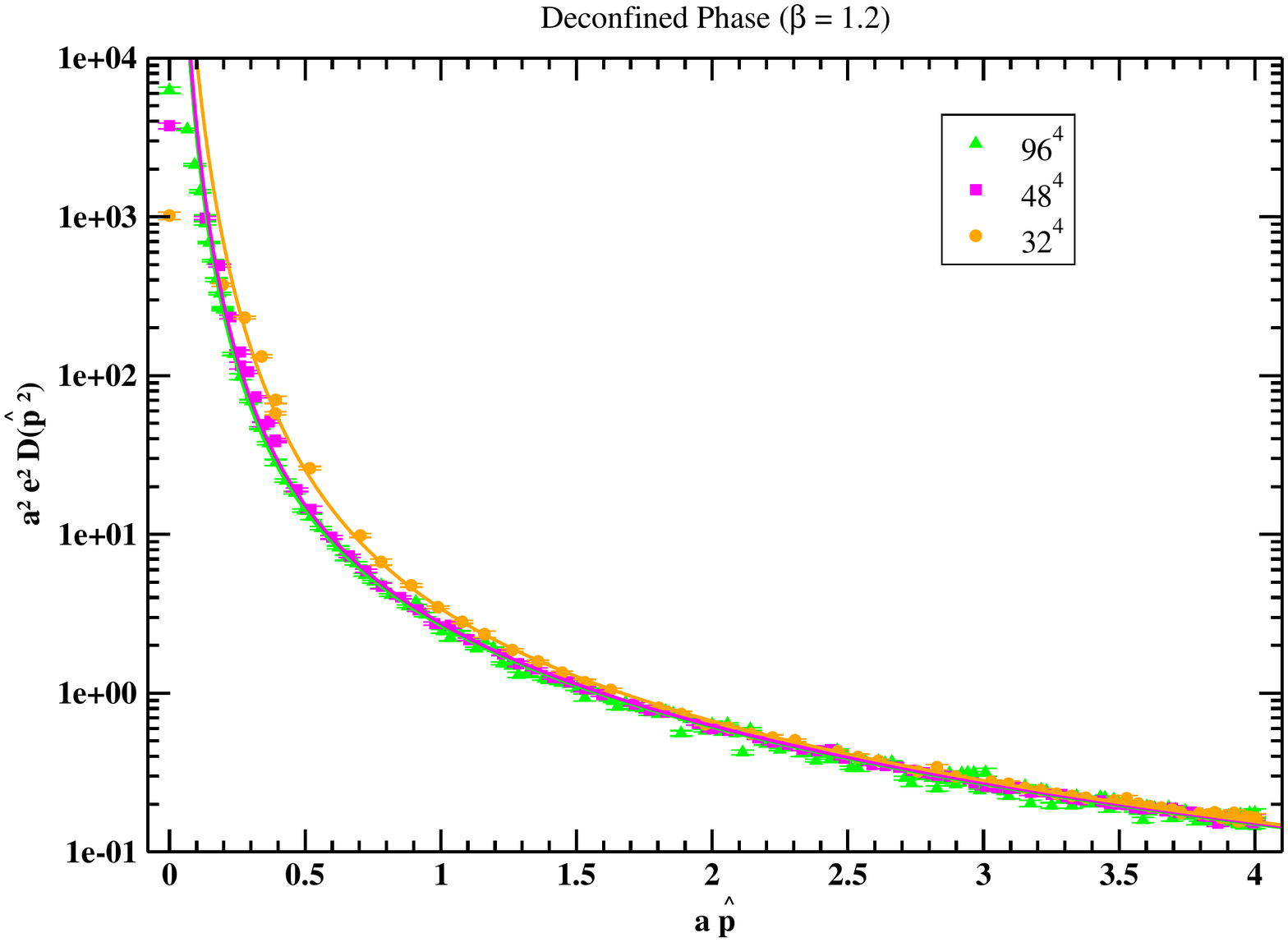}  \\
   \includegraphics[width=3.5in]{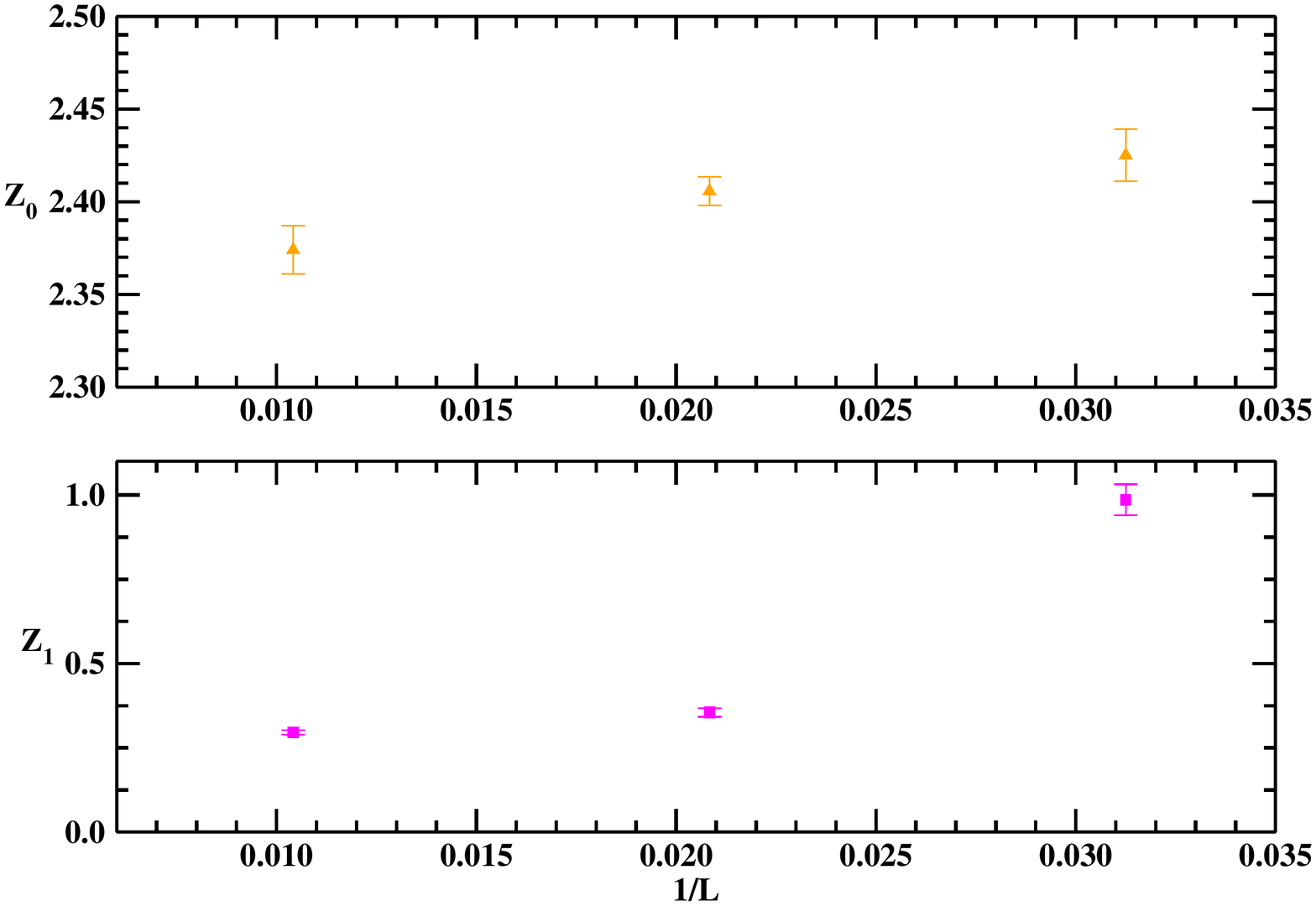}
   \caption{The Landau photon propagator in the deconfined phase and the fits to Eq. (\ref{Eq:prop_deconf}).}
   \label{fig:photon_deconf_fit}
\end{figure}

The data in Fig. \ref{fig:photon_deconf} suggest that  the photon propagator diverges as momentum approaches zero. If the data 
is to be associated with a free field theory, it should reproduce the behaviour of a free field propagator. However, in a finite volume Monte Carlo simulation  
deviations from the continuum free field theory are expected as the simulation is performed on a finite lattice. 
The approach to the continuum behaviour can be tested by fitting the lattice data to the functional form
\begin{equation}
  a^2  D(a^2 \hat{p}^2) = \frac{Z_0}{( a \,  \, \hat{p})^2 } + \frac{Z_1}{( a \,  \, \hat{p})^4 } \ .
  \label{Eq:prop_deconf}
\end{equation}  
If the theory reproduces a free field theory a $Z_1 \neq 0$ is a manifestation of finite volume effects and one expects $Z_1$ to become smaller as the lattice volume
is increased. The direct fit using the full range of momenta and taking into account the statistical errors of the lattice data
returns values of the $\chi^2/d.o.f. \gtrsim 12$. For the smallest volume, for $a \hat{p} \geqslant 0.6$ the fit has
a $\chi^2/d.o.f. = 1.74$ with $Z_0 = 2.425(14)$ and $Z_1 = 0.986(46)$. For the $48^4$ data and for $a \hat{p} \geqslant 0.4$ it follows that
$\chi^2/d.o.f. = 1.01$ with $Z_0 = 2.4057(76)$ and $Z_1 = 0.355(13)$. On the other hand for the largest lattice volume, due to large fluctuations that
are observed at  larger momenta, one can never achieve a reasonable $\chi^2/d.o.f.$ However, by doubling the statistical errors on the definition of the
$\chi^2$, the data becomes compatible with (\ref{Eq:prop_deconf}) for $a \hat{p} \geqslant 0.15$. In this case the fit has 
$\chi^2/d.o.f. = 1.93$ with $Z_0 = 2.374(13)$ and $Z_1 = 0.2958(68)$. Note that in all cases one has $Z_0 \approx 2.4$, while $Z_1$ decreases with
the lattice volume.\footnote{The data for $Z_0$ and $Z_1$ in Fig.  \ref{fig:photon_deconf_fit} suggest that $Z_0$ approaches its infinite
volume limit linearly. A linear extrapolation towards$L \rightarrow \infty$ results in $Z_0 = 2.352(10)$. On the other hand, $Z_1$ clearly does not
follow a linear behavior towards the infinite volume limit. Given that our simulations only considered three different volumes it is not possible to
estimate $Z_1 (L \rightarrow \infty)$.}.

The lattice data together with the fits can be seen in Fig. \ref{fig:photon_deconf_fit}. In general and for the 
corresponding fitting ranges, the curves overlap the Monte Carlo data. Further, the coefficient $Z_0 \approx 2.374(13)$, we are quoting the value of the fit
to the largest lattice volume, seems to be nearly independent of the volume.
The data in Fig. \ref{fig:photon_deconf_fit} suggest  that $Z_0$ is independent of $L$, while 
$Z_1$ is sensitive to $L$. Indeed this coefficient goes from $Z_1 = 0.986(46)$ for the smallest volume to 
$Z_1 = 0.2958(68)$ for the largest volume, which is about $1/3$ of the value for the smallest volume; note that the inverse of the ratio of the lattice sizes is 1/3.
This result for $Z_1$ suggests that the data for the propagator seems to converge to the propagator of a free field theory in the infinite volume limit. 
This statement has to be read with care due to the use of $2 \, \sigma$ in the definition of the minimising $\chi^2$ for the largest volume. 
That the fitting range does not start at the smallest non vanishing momentum for each volume is not surprising, as finite volume effects, that should
appear at the smallest momenta, are to be expected. 
We have also tried fitting the data with an almost free field propagator, i.e. assuming $D(p^2) = Z_0 / (p^2)^\alpha$
and leaving $Z_0$ and $\alpha$ as free parameters. The fits to this last functional form have the same problems as those mentioned before but it turns out that
$\alpha \approx 1$, i.e. the lattice data for the propagator follows closely the behaviour of a free field 
theory\footnote{In order to quantify the typical values of $\alpha$ let us report on its values given by fitting to the propagator data replacing
$\sigma$ by $ 2 \sigma$ in the definition of the minimising $\chi^2$. Demanding that the $\chi^2/d.o.f. \leqslant 2$, it follows
that for the smallest lattice volume the fitting range starts at $a \hat{p} = 0.6$ and has $\alpha = 1.095/10)$,
the fitting range for the $48^4$ data starts at $a \hat{p} = 0.4$ and has $\alpha = 1.0529(60)$, while for the
largest volume the fitting range starts at $a \hat{p} = 1$ and has $\alpha = 0.9977(74)$.}.

\begin{figure}[t] %  figure placement: here, top, bottom, or page
   \centering
   \includegraphics[width=3.5in]{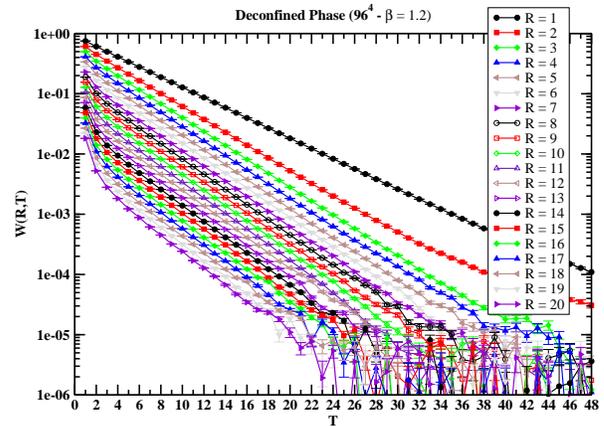}  \\
   \includegraphics[width=3.5in]{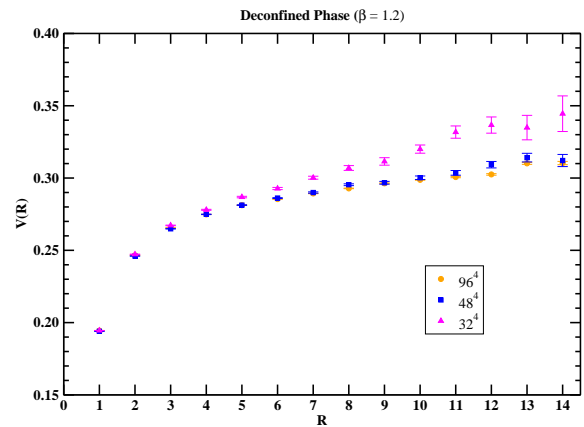}
   \caption{The Wilson loop (top) for $L = 96$ and the stattic potential (bottom) for all the lattice volumes in the deconfined phase.}
   \label{fig:photon_deconf_VR}
\end{figure}

\begin{figure}[t] %  figure placement: here, top, bottom, or page
   \centering
   \includegraphics[width=3.4in]{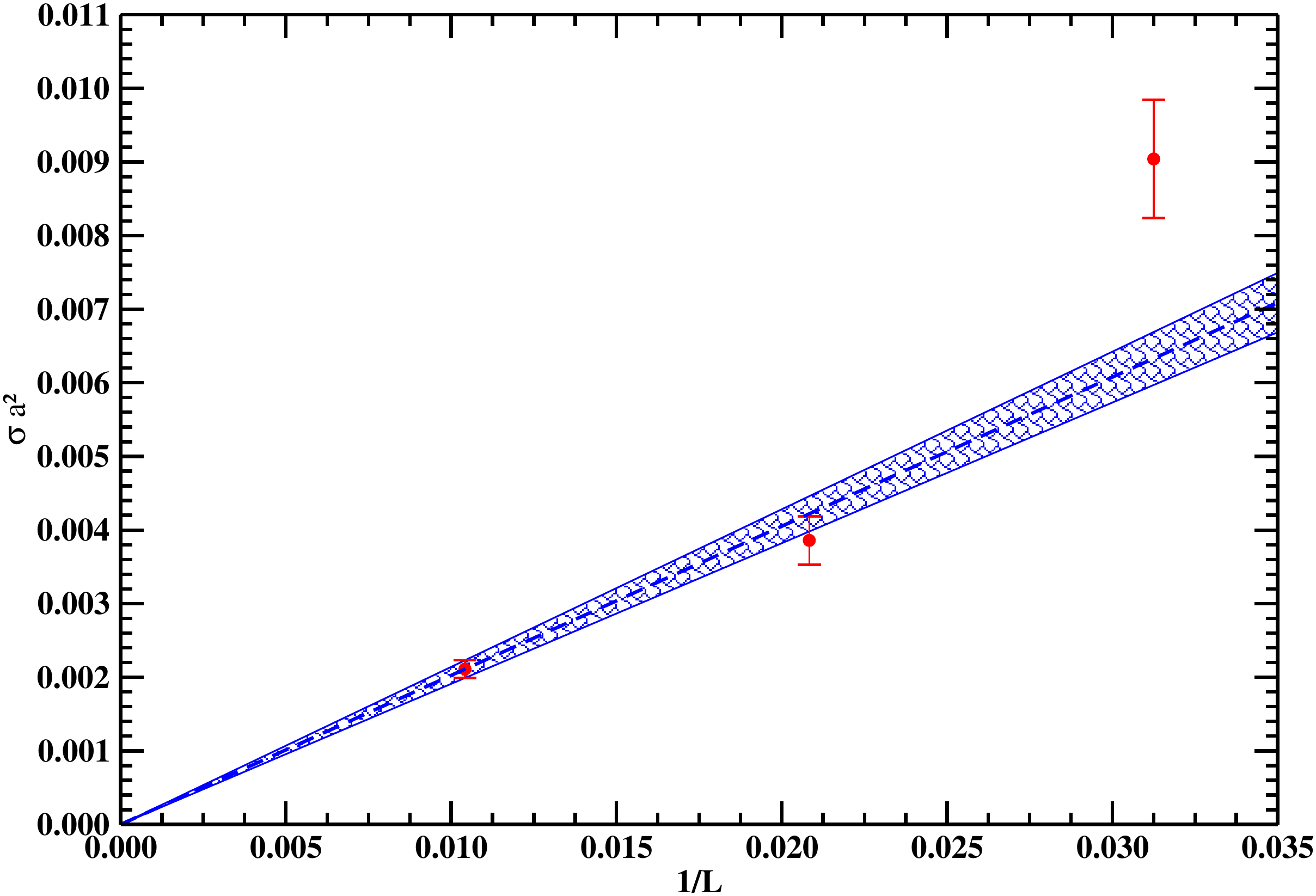}
   \caption{The string tension as a function of $1/L$. The lines are connect the origin where $\sigma \, a^2 = 0$ with the value found for the largest lattice volume. The 
                shaded region represents the one standard deviation on the result for $\sigma \, a^2$ from the largest volume.}
   \label{fig:photon_deconf_string}
\end{figure}

The above analysis suggests that the 
Monte Carlo propagator data almost reproduce a free field theory propagator. Let us check the results for the static potential, computed from Wilson loops
as was done for the confined phase. The Wilson loop for various values of $R$ is given in Fig. \ref{fig:photon_deconf_VR} for the largest lattice volume and it
looks rather different from the Wilson loop for the confined phase reported in Fig. \ref{fig:wilson_loop_conf}. If for the confined phase the slope increases with $R$,
for the deconfined phase the slope of the $\log W(R,T)$ seems to be the same for all $R$. 
Indeed, measuring $V(R)$ from the effective mass and taking its value for T = 9, one 
 gets the bottom plot of Fig. \ref{fig:photon_deconf_VR}. The large distance behaviour of $V(R)$ is sensitive to finite volume effects and 
the slope of $V(R)$ for large $R$ becomes smaller as $L$ is increased. In Fig. \ref{fig:photon_deconf_string} we show the string tension measured by fitting
$V(R)$ in the range $R = 9 - 12$ for the various volumes. The corresponding $\chi^2/d.o.f.$ for the various fits are always below 0.5.
The dashed blue line connects the origin with the result for the largest volume, while the shaded region takes into account the one standard deviation on 
$\sigma a^2$ for $L = 96$. Our results seems to be compatible with a vanishing string tension in the infinite volume limit. 

The short distance behaviour of $V(R)$ is difficult to understand from the computed Wilson loop directly.
The Monte Carlo data for the photon propagator is compatible with free field propagator behaviour at high $p$ and approaching $1/p^2$ as the volume is
increased and, therefore, one expects to have, in the infinite volume limit, $V(R) \propto 1/R$ at short distances. 
We have tried to  disentangle the short distance behaviour from the $V(R)$ Monte Carlo data but the results were inconclusive. 

Our simulations for the deconfined phase of compact QED
suggest that for the $\beta$ considered here, the finite volume effects are still not negligible even for a lattice volume as large as $96^4$.

%===================================================================================
%===================================================================================
\section{Summary and Conclusion \label{sec:fim}}

\begin{figure}[t] %  figure placement: here, top, bottom, or page
   \centering
   \includegraphics[width=3.4in]{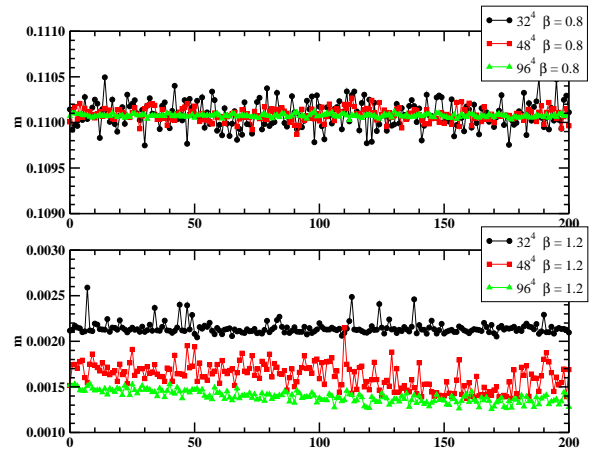}
   \caption{Average Dirac string density over the lattice as given by Eq.  (\ref{Eq:mean_dirac_string}) for the confined (top) and deconfined (bottom) phases 
                 for the thermalised Landau gauge configurations. The horizontal axis refers to the configuration number.}
   \label{fig:monopoles}
\end{figure}

In the current work the Landau gauge photon propagator is investigated for compact QED in the strong coupling (confining) and weak coupling (free field theory)
regimes and for various lattice volumes. By computing the static potential, our simulation confirms that at low $\beta$ the theory is confining and the behaviour
of the photon propagator in momentum space follows closely a Yukawa type of propagator, i.e. 4D compact QED has a mass gap.
Moreover, in the confining phase the theory develops a mass scale that makes the photon propagator finite in the full momentum range. 

For the deconfined phase, the photon propagator seems to approach a free field type of propagator as the infinite volume is approached. 
We have observed that the matching with a free field theory is not perfect with both the photon propagator and the static potential showing
some deviations from the expected behaviour, that we interpreted as being due to finite volume effects. Indeed, the deviations from a free field theory 
are reduced, in all the computed quantities, as the lattice volume is increased. Given that at low momenta the propagator of a free field theory
diverges and the lattice regularizes both the UV and IR divergences, in a sense the deviations from the free field theory results are not unexpected.

Comparing the confined and deconfined phase results, it seems that it is the generation of a mass gap that occurs for the confined phase that turns the 
propagator essentially independent of the lattice volume. This is a situation that is also seen in the simulations for QCD.

If the phase diagram for compact QED as a function of $\beta$ has two different phases, one can ask how can one distinguish them. 
According to \cite{Polyakov:1975rs} the appearance of the confining phase for the 3D theory is due to the presence of monopole configurations.
The monopoles are connected with the topology of the gauge group and they becomes irrelevant for the dynamics at large $\beta$ values. 
The 4D equivalent to the monopole configurations are Dirac strings that should be seen on a finer analysis of the gauge configurations. 
In Fig. \ref{fig:monopoles} we report on the average number of Dirac strings over the lattice, computed with the definitions (\ref{EqDirac_string0}) 
and (\ref{Eq:mean_dirac_string}), for all the lattices. 
The plots show that $m$ is independent of $L$    %.....in both phases, although the fluctuations for 
for the confined phase ($\beta = 0.8$), with $m$ having larger fluctuations in the smaller lattices. 
For the deconfined phase ($\beta = 1.2$) $m$ decreases with the lattice volume and the results suggest, once more, that 
in the simulations performed in the deconfined phase the infinite volume limit is not yet realised, despite the large volume considered.
%the smaller lattices are much larger, and that 
Furthermore, $m$ is about a factor of fifty larger in the confined phase when compared to its value in
the deconfined phase. This result suggests that, indeed, the Dirac strings are responsible for the confined phase in 4D compact QED in
agreement with the suggestion of  \cite{Polyakov:1975rs}. However, further studies are required to draw firm conclusions.

%=====================================================================================
%=====================================================================================
\section*{Acknowledgments}

This work was partly supported by the FCT – Funda\c{c}\~ao para a Ci\^encia e a Tecnologia, I.P., under Projects Nos. 
UIDB/04564/2020 and UIDP/04564/2020.
P. J. S. acknowledges financial support from FCT (Portugal) under Contract No. CEECIND/00488/2017.
The authors acknowledge the Laboratory for Advanced Computing at the University of Coimbra (\url{http://www.uc.pt/lca}) 
for providing access to the HPC resource Navigator.

%%%%%%%%%%%%%%%%%%%%%%%%%   Bibliography   %%%%%%%%%%%%

\end{document}